\documentclass[twocolumn,jcp,showpacs]{revtex4}
\usepackage{psfrag}
\usepackage{graphicx}
\usepackage{amsmath}
\usepackage{amssymb}

\begin{document}
\title{Non-equilibrium \emph{GW} Approach to Quantum Transport in
  Nano-scale Contacts}
\author{Kristian S. Thygesen$^1$}\author{Angel Rubio$^{1,2}$}
\affiliation{$^1$ Institut f{\"u}r Theoretische Physik, Freie
Universit{\"a}t Berlin, Arnimallee 14, D-14195 Berlin, Germany \\
$^2$Departamento de F\'\i sica de Materiales,
Facultad de Ciencias Qu\'\i micas, University of the Basque Country UPV/EHU, Centro Mixto, and European Theoretical Spectroscopy Facility (ETSF). E--20018 San Sebasti\'an, Basque Country, Spain}
\date{\today}

\begin{abstract}
  Correlation effects within the $GW$ approximation have been
  incorporated into the Keldysh non-equilibrium transport formalism.
  We show that $GW$ describes the Kondo effect and the
  zero-temperature transport properties of the Anderson model fairly
  well. Combining the $GW$ scheme with density functional theory and a
  Wannier function basis set, we illustrate the impact of correlations
  by computing the $I$-$V$ characteristics of a hydrogen molecule
  between two Pt chains. Our results indicate that self-consistency is
  fundamental for the calculated currents, but that it tends to
  wash out satellite structures in the spectral function.
\end{abstract}

\pacs{72.10.Bg,73.23.-b,73.63.Rt} \maketitle Electronic correlations
are responsible for important transport phenomena such as Coulomb
blockade and Kondo effects~\cite{goldhaber}, yet its significance for
transport in nano-scale structures is not well understood nor has it
been systematically studied. At present, the most popular approach to
\emph{ab intio} simulations of transport in nanocontacts combines a non-equilibrium Green's function
formalism with the single-particle Kohn-Sham (KS) scheme of
density functional theory (DFT). This approach works well for some
systems~\cite{sknielsen,djukic}, but in other cases it fails to
reproduce experimental data~\cite{stokbro03} indicating the need for computational transport schemes beyond the DFT level~\cite{delaney_prl,ferretti_prl,darancet}.

A reliable description of transport through a molecular junction requires
first of all a reliable description of the electronic structure of the
molecule itself, i.e. its electron addition- and removal energies. It
is well known that the $GW$ self-energy method yields quasiparticle
properties of molecules~\cite{stan06,niehaus05} and solids~\cite{hybertsen86,rmp} in good agreement with experiment
improving drastically the DFT band structures.
In view of this it seems tempting to extend the use of the $GW$
approximation to transport calculations. It is clear, however, that this should not be implemented by shifting the molecular energy levels to their $GW$ positions prior to
coupling. The reason is, that when a confined interacting system is connected to external
(non-interacting) leads, the electrons in the confined region become
correlated with those in the leads. To capture these correlations,
which are the origin of important many-body phenomena such as the
Kondo effect, it is crucial that the self-energy be evaluated in the presence of coupling to
the leads. 

Traditionally, correlation effects in transport have been studied on the basis of the
Anderson- and Kondo models by a variety of numerical and analytical
techniques. Many of these techniques are, however, quite specific
to the considered models and lack the generality needed to be
combined with first-principles methods. 

In this paper we combine the $GW$ approximation with the non-equilibrium Keldysh formalism to obtain a practical scheme for correlated quantum transport. We study the
Anderson model out of equilibrium, and calculate the $I$-$V$
characteristics of a molecular hydrogen contact using a Wannier
function (WF) basis set. In
both applications we emphasize the difference between self-consistent
and non self-consistent evaluations of the $GW$ self-energy.

As a general model of a quantum conductor we consider a central region
($C$) connected to left ($L$) and right ($R$) leads. The leads are kept at chemical potentials  $\mu_L$ and $\mu_R$, respectively. We construct the matrix $h_{ij}=\langle \phi_i|\hat h_s|\phi_j\rangle$, where $\hat h_s=-\frac{1}{2}\nabla^2+v_h+v_{xc}+v_{ext}$ is the KS Hamiltonian of the combined $L$-$C$-$R$ system in equilibrium, and $\{\phi_i\}$ is a corresponding set of maximally localized, partly occupied WFs~\cite{WFprl}.
Assuming that correlation effects as well as charge redistributions induced by the bias voltage are significant only inside $C$, we describe the leads and the coupling to the central region by $h$. Using standard methods~\cite{thygesen_chem_phys05} we evaluate the coupling self-energies, $\Sigma_{\alpha}(\omega)=h_{C \alpha}g_{\alpha \alpha}(\omega) h_{\alpha C}$, where $g_{\alpha\alpha}$ is the GF of the uncoupled lead $\alpha=L,R$. The interactions inside the central region are described by $\hat V_{int}=\frac{1}{2}\sum_{ijkl \in C,\sigma
  \sigma'}V_{ij,kl}c^{\dagger}_{i\sigma}c^{\dagger}_{j\sigma'}c_{l\sigma'}c_{k\sigma}$,
with the Coulomb matrix elements $V_{ij,kl}=\int \int \text{d}\bold r
\text{d}\bold r' \phi_i(\bold r)^*\phi_j(\bold r')^* \phi_k(\bold
r)\phi_l(\bold r') /|\bold r - \bold r'|$.
In practice we use an effective interaction which
only involves a subset of the Coulomb matrix elements, see later. We
include matrix elements of the form $V_{ij,ij}$ and $V_{ij,ji}$ in the
calculation of the correlation part of the $GW$ self-energy while also
terms of the form $V_{ii,jj}$ and $V_{ii,ij}$ are included in the
Hartree and exchange self-energies~\cite{comment}. By allowing the
effective interaction to be spin-dependent we avoid the
self-interaction normally present in the $GW$ correlation.

We calculate the retarded and lesser Green's functions of the central
region from~\cite{haug_jauho,extraterm}
\begin{eqnarray}\label{eq.gr}
G^r&=&[\omega+i\eta-h_{CC}+(v_{xc})_{CC}-\Delta v_h-\Sigma_{tot}^r]^{-1}\\ \label{eq.gl}
G^<&=&G^r\Sigma_{tot}^< G^a+2\eta i f_C G^r G^a,
\end{eqnarray}
where $\eta$ is a positive infinitesimal and $\Sigma_{tot}=\Sigma_L+\Sigma_R+\Sigma$ is the sum of the coupling self-energies and the exchange-correlation part of the interaction
self-energy. The term $\Delta v_h= \Sigma_h^r[G]-\Sigma^r_h[G_{\text{DFT}}^{eq}]$, is the correction to the equilibrium (DFT) Hartree potential introduced by the correlations and the finite bias. The DFT xc-potential is subtracted from $h_{CC}$ to avoid double counting. In the last term of Eq.~(\ref{eq.gl}), $f_C$ denotes the initial Fermi-Dirac distribution of the central region before coupling to the leads. Notice that this term can become
significant at energies where $\text{Im}\Sigma_{tot}^r$ is comparable to
$\eta$~\cite{extraterm}. While this is not expected to occur in the
interacting case (due to life-time broadening by $\Sigma$), it will
happen in the non-interacting case whenever bound states are present in
$C$. We note, that in
Eqs. (\ref{eq.gr}) and (\ref{eq.gl}) we have specialized to the long
time limit where we assume that the system reaches a steady state in which the
Green's functions depend only on the time difference $t=t_2-t_1$ and
thus can be represented by a single time/frequency variable.\cite{stefanucci04}  

The symmetrized current, $I=(I_L+I_R)/2$, where $I_{L(R)}$ is the
current in the left (right) lead, is given by~\cite{meir_wingreen}
\begin{equation}\nonumber
I=\frac{i}{4\pi}\int \text{Tr}[(\Gamma_L-\Gamma_R)G^{<}+(f_L\Gamma_L-f_R\Gamma_R)(G^r-G^a)]\text{d}\omega
\end{equation}
where $\Gamma_{L(R)}=i[\Sigma^r_{L(R)}-\Sigma^a_{L(R)}]$ is the coupling strength and the trace is taken over basis functions in the central region.

Within the \emph{GW} approximation $\Sigma$ is written as a product of
the Green's function, $G$, and the screened interaction, $W$,
calculated in the random phase approximation (RPA). Out of equilibrium this
holds true on the Keldysh contour, and the relevant equations in real
time follow from the Langreth conversion
rules~\cite{haug_jauho}. Absorbing the spin index into the orbital index we define
the effective interaction $\hat
V_{\text{eff}}=\frac{1}{2}\sum_{ij}\tilde
V_{ij}c^{\dagger}_ic^{\dagger}_{j}c_{j}c_{i}$, where $\tilde
V_{ij}=V_{ij,ij}-\delta_{\sigma
  \sigma'}V_{ij,ji}$. $\hat V_{\text{eff}}$ resembles the real
space interaction with the important difference that $\tilde V_{ij}$ is spin-dependent and
$V_{ii}=0$. Self-interaction is thus automatically excluded to
all orders in $\hat V_{\text{eff}}$. The retarded and lesser $GW$ self-energies
become (on the time axis),
\begin{eqnarray}
\Sigma^r_{ij}(t)&=&iG^{r}_{ij}(t)W^{>}_{ij}(t)+iG^{<}_{ij}(t)W^{r}_{ij}(t)\\
\Sigma^<_{ij}(t)&=&iG^{<}_{ij}(t)W^{<}_{ij}(t).
\end{eqnarray}
The screened interaction is given by (in frequency space),
\begin{eqnarray}\label{eq.screen}
W^{r}(\omega)&=&\tilde V[I-P^{r}(\omega)\tilde V]^{-1}\\
W^{</>}(\omega)&=&W^{r}(\omega)P^{</>}(\omega)W^{a}(\omega),
\end{eqnarray}
where all quantities are matrices in the central region indices.
Finally the irreducible polarization becomes
\begin{eqnarray}
P^{r}_{ij}(t)&=&-iG^{r}_{ij}(t)G^{<}_{ji}(-t)-iG^{<}_{ij}(t)G^{a}_{ji}(-t)\\
P^{</>}_{ij}(t)&=&-iG^{</>}_{ij}(t)G^{>/<}_{ji}(-t).
\end{eqnarray}

In principle the \emph{GW} method implies a self-consistent problem,
i.e.  the $G$ obtained from Eqs.~(\ref{eq.gr}) and (\ref{eq.gl})
should equal the $G$ used to produce $\Sigma$. In this way
all important conservation laws are fulfilled~\cite{ulf}. However, due
to the large computational demands $GW$ band structure
calculations usually apply a, quite successful, non self-consistent $G_0W_0$ approach 
using on the $G_0$ obtained from a KS calculation~\cite{hybertsen86}. We mention here that the
non-equilibrium $GW$ approximation has previously been used in the
study of semiconductors in high-intensity laser
fields~\cite{haug_jauho,louie04}.

We represent all quantities on a uniform real time/frequency grid using the Fast Fourier transform to switch between the two representations to 
avoid time consuming convolutions. 
The calculation of the WFs, the matrix elements $h_{ij}$, and the
coupling self-energies $\Sigma_{\alpha}$ is described in detail in
Ref.~\onlinecite{thygesen_chem_phys05}.

Before turning to the Wannier-$GW$ calculations we apply the method
to the Anderson impurity model. Although $GW$ is not expected to be accurate
for strongly correlated systems, the simplicity of the Anderson model
makes it ideal for illustrating the properties and limitations of the
non-equilibrium $GW$ approximation. We thus consider a central site of
energy $\varepsilon_c$ coupled to one-dimensional leads with on-site
energy $\varepsilon_0=0$ and nearest-neighbor hopping $t_0=10$.  The
hopping to the central site is $t_c=1.8$ giving $\Gamma \equiv
\Gamma_L(0)=\Gamma_R(0)\approx 0.65$. Double occupation of the central
site costs a charging energy of $U=4$, and the spin-dependent
interaction entering Eq. (\ref{eq.screen}) is $\tilde V_{\sigma\sigma'}=U(1-\delta_{\sigma \sigma'})$. We assume
half-filled bands, i.e. $E_F=\varepsilon_0$, and measure all energies
relative to $E_F$.

We consider three different approximations: (i) Self-consistent Hartree, which
is equivalent to self-consistent Hartree-Fock (HF) since the exchange
term vanishes.  (ii) $G_0W_0$ with the self-consistent HF
Green's funtion as $G_0$. (iii) Fully self-consistent $GW$. All calculations are non-magnetic, i.e. $G_{\uparrow}=G_{\downarrow}=G$. 

\begin{figure}[!h]
\includegraphics[width=0.8\linewidth]{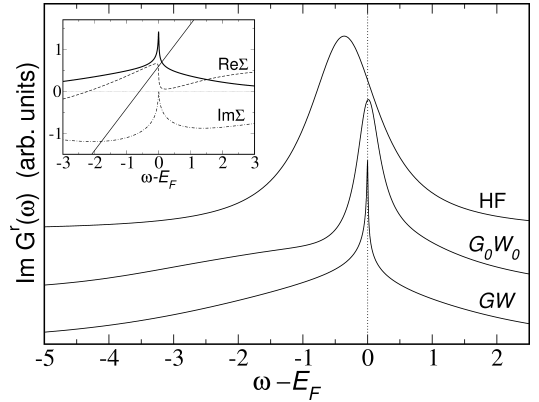}
\caption{\label{fig1} Spectral function at the central
  site, $\varepsilon_c=-3$, of the Anderson model calculated in three different ways (see text). The curves have been vertically
  offset for clarity. Inset: real and imaginary parts of the $GW$
  self-energy together with the line $\omega-\varepsilon_{\text{\emph{HF}}}$. The
  steep shape of $\text{Re}\Sigma^r$ around $E_F$ pins the position of
  the quasi-particle peak.}
\end{figure}

In Fig.~\ref{fig1} we show the equilibrium spectral function at the
central site, $\text{Im}G^r(\omega)$, for $\varepsilon_c=-3$.
The HF solution
shows a single peak at $\varepsilon_{\text{HF}}=\varepsilon_c+U\langle \hat
n_{\sigma} \rangle$ with a full width at half maximum given by
$2\Gamma$.  This behavior is representative for any mean-field
description, including the KS scheme. The inclusion of dynamic
correlations leads to qualitative changes in the spectral peak which
moves close to the chemical potential~\cite{chempot} and narrows down
from $2\Gamma$ to $0.63$ $(0.28)$ in the case of $G_0 W_0$ ($GW$).
This change is a signature of the Kondo effect: For $\Gamma-U<
\varepsilon_c < -\Gamma$ (the so-called Kondo regime), the
correlated groundstate is a singlet with a finite amplitude for the
central site being empty.  At $T=0$ this leads to the formation of a
spectral peak at the chemical potential with a width given by the
Kondo temperature, $T_K=\frac{1}{2}(2\Gamma
U)^{1/2}\exp[\pi\varepsilon_c(\varepsilon_c+U)/2\Gamma U]$. For our
choice of parameters $T_K=0.19$, which is in fair agreement with the
$GW$ result, and about three times smaller than the $G_0 W_0 $ result. 

From the inset of Fig.~\ref{fig1} it
can be seen that the Kondo peak gets pinned to $E_F$ due to the
steep shape of $\text{Re}\Sigma^r$ in this region, and that its
reduced width, as compared to $2\Gamma$, is a consequence of the steep
drop in $\text{Im}\Sigma^r$ away from the point
$\text{Im}\Sigma^r(\omega=E_F)=0$. The atomic levels which should be
seen at $\varepsilon_c$ and $\varepsilon_c+U$ appear as shoulders in the $G_0
W_0$ spectrum, but for larger values of $U/\Gamma$ they become more
pronounced as satellites (side-bands) of the main quasiparticle peak although their
positions are somewhat shifted towards $E_F$. In contrast the self-consistent
$GW$ fails to capture the side-bands. These findings
agree well with previous results obtained with the
fluctuation-exchange approximation~\cite{white92}, and with $GW$
studies of the homogeneous electron gas~\cite{holm_barth}. We mention
that in self-consistent second-order perturbation theory, the pinning
of the main spectral peak to $E_F$ is less pronounced than in $GW$ and
its width is significantly overestimated, showing as expected that the
higher order RPA diagrams enhance the strong correlation features.

\begin{figure}[!b]
\includegraphics[width=0.63\linewidth,angle=270]{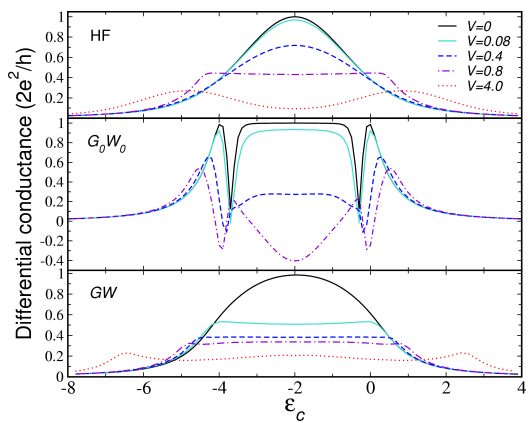}
\caption[cap.wavefct]{\label{fig2} (color online). Differential
  conductance in the Anderson impurity model as a
  function of the central site energy, $\varepsilon_c$, for different
  applied biases. The negative differential conductance seen in the
  middle panel is an artifact of the non-conserving nature of the
  $G_0W_0$ approximation.}
\end{figure}

In Fig.~\ref{fig2} we show the zero-temperature differential
conductance under a symmetric bias, $\mu_{L/R}=\pm V/2$, as a function
of $\varepsilon_c$. For $V=0$ there is only little difference between
the HF and $GW$ results which both shows a broad conductance peak
reaching the unitary limit at the symmetric point
$\varepsilon_c=-U/2$. The physical origin of the conductance trace is,
however, very different in the two cases: While the HF result is
produced by off-resonant transport through a broad spectral peak
moving rigidly through the Fermi level, the $GW$ result is due to
transport through a narrow Kondo peak which is always on resonance
(for $\varepsilon_c$ in the Kondo regime). This difference is brought
out clearly as $V$ is increased: for $V\ll \Gamma$ the bias has little
effect on the HF conductance while the $GW$ conductance drops
dramatically already at biases comparable to $T_K$ due to suppression
of the Kondo resonance. We note that we do not observe a
splitting of the $GW$ Kondo resonance at finite $V$~\cite{mwl93}. The
side peaks in the $dI/dV$ correspond to the central level crossing the
chemical potentials. The $G_0W_0$ conductance is markedly different
from the HF and $GW$ results. At low bias there are unphysical dips in
the conductance curve, and as $V$ is raised the $dI/dV$ becomes
even more unreasonable showing strong negative differential
conductance (for this reason we have omitted the $V=4$ curve). This
unphysical behavior of the $G_0 W_0$ is a result of its non-conserving
nature, and it underlines the necessity
of using a conserving approximation such as the self-consistent $GW$
for transport calculations. Within $G_0 W_0$ the
violation of current conservation, $(I_L-I_R)/I$, increases
with $V$ reaching $20\%$ at $V=0.8$ for certain values of $\varepsilon_c$, while it is always negligible in our $GW$ calculations~\cite{conserving}.

\begin{figure}
\includegraphics[width=0.75\linewidth]{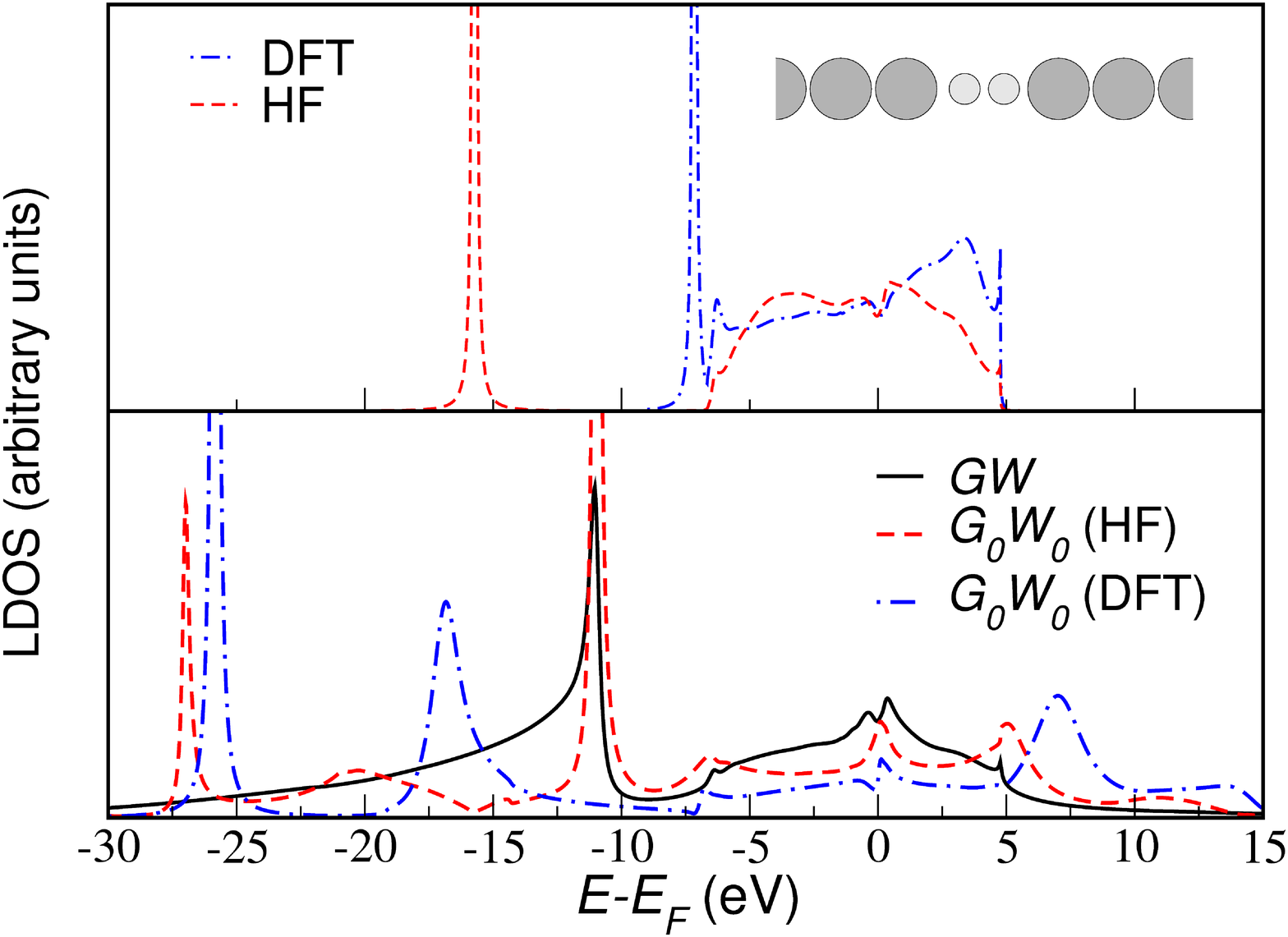}
\caption[cap.wavefct]{\label{fig3} (color online). Local density of states at one of the H orbitals of the
  Pt-H-H-Pt contact shown in the inset.}
\end{figure}

As an illustration of the Wannier-$GW$ scheme we consider a
molecular hydrogen bridge between infinite atomic Pt chains, see inset
of Fig.~\ref{fig3}.  Experimentally, the conductance of the hydrogen
contact is found to be close to the conductance quantum, $2e^2/h$, and
this value has been reproduced by DFT calculations.~\cite{djukic}
Using the plane-wave pseudopotential code Dacapo, we perform DFT calculations for an infinite Pt wire as well as a supercell containing the hydrogen molecule with six Pt atoms on each side. The WFs and KS Hamiltonian of the full system in equilibrium are obtained by combining the two calculations, see Ref. \onlinecite{thygesen_chem_phys05} for more details.
For the transport calculation we use a central region consisting of the two $s$-like WFs
of the hydrogen. Due to the smallness of the matrix elements
coupling the two Pt chains across the molecule this suffices to converge the elastic transmission function. Ultimately, the
dependence of the $GW$ results on the size of $C$ as well as on the
basis set should also be checked.~\cite{EPAPS}. 

In the upper panel of Fig.~\ref{fig3} we show the local density of
states (LDOS) at one of the two H orbitals as calculated within DFT using the PW91 xc-functional, as well as self-consistent HF (in the central region).
In DFT the $\text{H}_2$ bonding state
is a bound state at $-7.0$~eV relative to $E_F$, while the
anti-bonding state lies at $0.4$~eV and is strongly
broadened by coupling to the Pt. Moving from DFT to HF the bonding state is
shifted down by $\sim 8$~eV because for occupied states the exchange potential is more negative
than the DFT xc-potential. The same effect tends
to drive the half-filled anti-bonding state down but in this case the
resulting increase in the Hartree potential (about 4 eV) stops it just
below $E_F$.

In the lower panel of Fig.~\ref{fig3} we show the LDOS calculated in
$GW$ as well as $G_0W_0$ starting from either DFT or HF, i.e. $G_0$ is
either $G_{\text{DFT}}$ or $G_{\text{HF}}$. The large deviation
between the two $G_0W_0$ results is not surprising given the large
difference between $G_{\text{DFT}}$ and $G_{\text{HF}}$. Focusing on
the bonding state, the $G_0W_0$ quasiparticle (QP) energies lie at
$-26$~eV and $-11$~eV for DFT and HF, respectively. Two effects are responsible for
this difference: We have
$\varepsilon_{\text{QP}}(G_0)=\varepsilon_{\text{HF}}(G_0)+\varepsilon_{\text{\emph{corr}}}(G_0)$,
where $\varepsilon_{\text{\emph{corr}}}$ is determined by the intersection of the line
$\varepsilon-\varepsilon_{\text{HF}}$ with the real part of the
correlation self-energy, $\text{Re}\Sigma^r_{\text{\emph{corr}}}$. Now,
$\varepsilon_{\text{HF}}(G_{\text{HF}})$ is already $\sim 5$~eV larger
than $\varepsilon_{\text{HF}}(G_{\text{DFT}})$ due, mainly, to the
mentioned increase in Hartree energy. Secondly, it turns out that
$\Sigma^r_{\text{\emph{corr}}}(G_{\text{HF}})$ is roughly $\Sigma^r_{\text{\emph{corr}}}(G_{\text{DFT}})$
shifted down by $\sim 9$~eV (note that this corresponds to the
difference between the delta peaks in $G_{\text{HF}}$ and
$G_{\text{DFT}}$), leading to $\varepsilon_{\text{\emph{corr}}}(G_{\text{DFT}})\approx -6$~eV
and $\varepsilon_{\text{\emph{corr}}}(G_{\text{HF}})\approx 5$~eV~\cite{EPAPS}. We
are aware that part of this difference could be due to the
limited size of the basis.

We notice, that the LDOS results of Fig.~\ref{fig3} can be largely reproduced by including only the
second-order $GW$ diagram in the self-energy. Thus the higher-order RPA
diagrams are less important in this case.

\begin{figure}
\includegraphics[width=0.49\linewidth,angle=270]{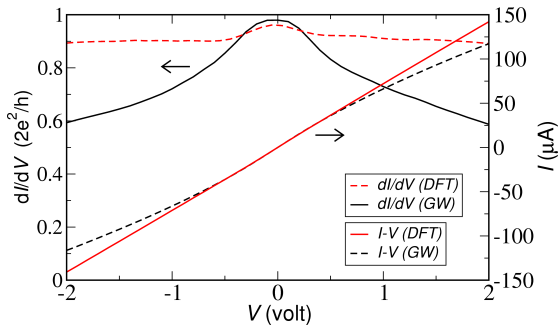}
\caption[cap.wavefct]{\label{fig4} (color online). $I$-$V$ and $dI/dV$
  for the hydrogen contact as calculated in DFT(PW91) and self-consistent
  $GW$. $V$ is the source-drain bias voltage.}
\end{figure}

The linear-response conductance has been calculated by applying a small bias of $10$ mV. All the self-consistent calculations, i.e. DFT, HF, and $GW$, yield a
conductance within $10\%$ of the experimental value of
$2e^2/h$. The same holds for \mbox{$G_0W_0$(HF)}, however, this is somewhat arbitrary as the
\mbox{$G_0W_0$(DFT)} conductance is only $0.4(2e^2/h)$. In
Fig.~\ref{fig4} we show the fully self-consistent $I$-$V$
characteristics for DFT and $GW$. The DFT conductance is nearly
constant over the bias range (like the HF, not shown). In contrast the $GW$ conductance falls off at higher bias due to incoherent scattering described by $Im\Sigma_{corr}$. Since
$\text{Im}\Sigma_{\text{\emph{corr}}}(E_F)$ vanishes in equilibrium,
the finite-bias conductance suppression is a direct result of the non-equilibrium
treatment of correlations.

In conclusion, we have presented a non-equilibrium $GW$ implementation 
that can be combined with DFT and a localized basis set
to model correlated electron transport in nanostructures.
Results for the Anderson model and a Pt-H-H-Pt
nano-contact indicate that self-consistency  is crucial for $GW$ 
transport calculations. 

We thank E. K. U. Gross for stimulating discussions 
during our stay in his group. The authors were partly
supported by the EC 6th Framework NoE NANOQUANTA and SANES project. K. T. acknowledges support from the Danish
Natural Science Research Council.


\bibliographystyle{apsrev}

\begin{thebibliography}{24}
\bibitem{comment} With our localized basis set we have found
that this is sufficient to reproduce Hartree and exchange energies of
most molecular orbitals to within $10\%$.
\bibitem{extraterm} The last term in Eq.~(\ref{eq.gl}) is usually
      written as $(1+G^r\Sigma_{tot}^r)G_0^<(1+\Sigma_{tot}^a G^a)$,
      where $G_0$ is the Green's function of
      $H_C+V_H$~\cite{haug_jauho}. Using that
      $G_0^<(\omega)=-f_C(\omega)(G_0^r-G_0^a)=2f_C(\omega) \eta i$,
      the equivalence of the two forms is readily shown.
\bibitem{chempot} In the Kondo regime, the
  $G_0W_0$ spectral peak can appear more or less close to $E_F$
  depending on the values of $\Gamma$ and $U$, whereas the $GW$ 
  peak is always situated very close to $E_F$ independent of $U$ and
  $\Gamma$.
\bibitem{conserving}
  The proof that $I_L=I_R$ for the present self-consistent $GW$
  calculations will be given elsewhere.
\end{thebibliography}

\end{document}